\documentclass[conference]{IEEEtran}
\IEEEoverridecommandlockouts
\usepackage{cite}
\usepackage{amsmath,amssymb,amsfonts}
\usepackage{algorithmic}
\usepackage{graphicx}
\usepackage{textcomp}
\usepackage{tikz}
\usepackage{xcolor}
\usepackage{multirow}
\usepackage{hyperref}
\def\BibTeX{{\rm B\kern-.05em{\sc i\kern-.025em b}\kern-.08em
    T\kern-.1667em\lower.7ex\hbox{E}\kern-.125emX}}
\begin{document}

\title{LProtector: an LLM-driven Vulnerability Detection System\\
}

\author{
\IEEEauthorblockN{Ze Sheng}
\IEEEauthorblockA{\textit{Computer and Information Science} \\
\textit{University of Pennsylvania}\\
Philadelphia, USA \\
zesheng.upenn@gmail.com}
\and
\IEEEauthorblockN{Fenghua Wu}
\IEEEauthorblockA{\textit{Integrated Circuit Science \& Engineering} \\
\textit{University of Electronic Science \&} \\
\textit{Technology of China}\\
Chengdu, China \\
qicai@std.uestc.edu.cn}
\and
\IEEEauthorblockN{Xiangwu Zuo}
\IEEEauthorblockA{\textit{Computer Science \& Engineering} \\
\textit{Texas A\&M University}\\
College Station, USA \\
dkflame@tamu.edu}
\and
\IEEEauthorblockN{Chao Li}
\IEEEauthorblockA{\textit{Graduate School of Arts and Sciences} \\
\textit{Georgetown University}\\
Washington, D.C., USA \\
cl1486@georgetown.edu}
\and
\IEEEauthorblockN{Yuxin Qiao}
\IEEEauthorblockA{\textit{Computer Information Technology} \\
\textit{Northern Arizona University}\\
Flagstaff, USA \\
yq83@nau.edu}
\and
\IEEEauthorblockN{Lei Hang}
\IEEEauthorblockA{\textit{LG Energy Solution (Nanjing) Co., Ltd} \\
Nanjing, China \\
leihang988@gmail.com}
}

\maketitle
\maketitle
\begin{tikzpicture}[remember picture, overlay]
    \node[anchor=south west, xshift=1.6cm, yshift=1cm] at (current page.south west) {
        \begin{minipage}{\textwidth} 
            \scriptsize
            \rule{\textwidth}{0.5pt} \\ 
            \textbf{This is a preprint version of the article. The final version will be published in the proceedings of the IEEE conference.}
        \end{minipage}
    };
\end{tikzpicture}

\begin{abstract}
The security issues of large-scale software systems and frameworks have become increasingly severe with the development of technology. As complexity of software grows, vulnerabilities are becoming more challenging to detect. Although traditional machine learning methods have been applied in cybersecurity for a long time, there has been no significant breakthrough until now. With the recent rise of large language models (LLMs), a turning point seems to have arrived. The powerful code comprehension and generation capabilities of LLMs make fully automated vulnerability detection systems a possibility. This paper presents LProtector, an automated vulnerability detection system for C/C++ codebases based on GPT-4o and Retrieval-Augmented Generation (RAG). LProtector performs binary classification to identify vulnerabilities in target codebases. To evaluate its effectiveness, we conducted experiments on the Big-Vul dataset. Results show that LProtector outperforms two state-of-the-art baselines in terms of F1 score, demonstrating the potential of integrating LLMs with vulnerability detection. 
\end{abstract}


\section{Introduction}
AI has significantly progressed in various defect detection areas over a long period of time. An example is when Wang et al.~\cite{b1} utilized AI for identifying medical problems, while Wu and collaborators~\cite{b2},~\cite{b3} employed it for detecting Electromigration problems. Defects in software, also referred to as software vulnerabilities, pose a significant challenge in the cybersecurity field. Xu et al.~\cite{b4} demonstrate the various ways in which harm and software loss can arise from these vulnerabilities. Several restrictions are associated with conventional detection methods. Yao et al.~\cite{b5} and Li et al.~\cite{b6} discovered that Automated Program Repair (APR) depends on predetermined patterns or strategies to create patches. Nevertheless, the patches do not meet the expected quality standards. Code generated by APR may successfully meet certain test cases, but it could still struggle to address the underlying issue, leading to ineffective outcomes in different situations.

Klees et al., Kai et al., and Han et al.~\cite{b7},~\cite{b8},~\cite{b9} propose that while fuzzing tests are good at uncovering memory management errors, they are not as efficient at identifying intricate problems like race conditions and privilege escalation. Understanding the goals and procedures of a program is crucial in identifying logical errors, as depending only on fuzzing is insufficient for detecting and correcting them. Drawbacks are also linked to Static Analysis Tools (SAT). Pereia~\cite{b10} highlighted that SAT tools generate many incorrect results because they do not take into account dynamic factors such as variable value changes in real-time. They frequently sound the alarm for issues that are not real. An instance could be discovering a possible buffer overflow in a code path that is not used. Additionally, Liu et al.~\cite{b11} pointed out the difficulties SAT faces when handling intricate dynamic behaviors, such as dynamic memory allocation or conditional branches, leading to constraints in identifying dynamic vulnerabilities such as race conditions.These limitations highlight the need for more advanced methods. Similar innovations are needed for road damage detection. Han-Cheng Dan et al.~\cite{b12}successfully improved detection accuracy and efficiency using the enhanced YOLOv7 algorithm, demonstrating the advantages of improved model strategies.

In contrast, LLMs have powerful code generation and understanding capabilities~\cite{b13},~\cite{b14},~\cite{b15}, along with a rich knowledge base and strong generalization ability~\cite{b16},~\cite{b17},~\cite{b18}. For instance, Tan et al. demonstrated that neural networks could effectively convert textual descriptions into 3D models using encoder-decoder architectures, showcasing the potential of LLMs to handle multi-modal tasks across various domains ~\cite{b19}. Zhang et al. finds the black-box LLMs like GPT-4o can label textual datasets with a quality that surpasses that of skilled human annotators, which offers a new perspective for automated software defect detection, as LLMs can achieve more efficient training and labeling when handling large volumes of unlabeled data~\cite{b20}.  Thus, we selected the GPT-4o, which is currently one of the most capable models, as the AI agent for LProtector. This ensures good robustness even in systems with strong interference~\cite{b21}. We used the Big-Vul dataset~\cite{b22} to evaluate how well LProtector works by measuring its performance against VulDeePecker~\cite{b23} and Reveal~\cite{b24}. To enhance LProtector's cybersecurity knowledge, we used RAG methods to pick 500 random instances of CWE/CVE from the Big-Vul dataset and stored them in a vector database.

\section{Proposed Methodology}

\subsection{Architecture}
\begin{figure}[ht!]
\centerline{\includegraphics{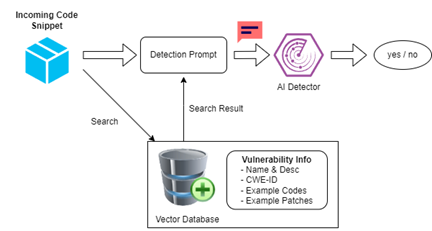}}
\caption{Architecture of LProtector}
\label{fig:arch}
\end{figure}

Fig.~\ref{fig:arch} shows the architecture diagram of LProtector. In LProtector, all input data is code snippets from the Big-Vul Dataset.

For data preprocessing, Pandas is used to extract CWE-ID, code description, vulnerability name and code snippets in the dataset. Then the metadata will be stored into a .json file which will be later embedded into vectors by OpenAI Embedding Algorithm. In this paper, ChromaDB is the vector database.

The input data is first mapped into vectors through word embedding~\cite{b25}. Deep reinforcement learning shows similar potential for complex decision-making, such as in autonomous navigation~\cite{b26}. Then, using an Augmented Prompt, the vectors are queried against the database to obtain the relevant results.Next, with Chain of Thought (CoT) prompt engineering method~\cite{b27},~\cite{b28}, the AI agent attempts to determine whether the code block contains a vulnerability and performs binary classification, where 1 indicates "yes" and 0 indicates "no".

\subsection{Retrieval-Augmented Generation}
RAG was proposed for a long time, but with the rise of LLMs, their powerful understanding capability can significantly improve the accuracy and precision of RAG's retrieval results~\cite{b29},~\cite{b30}. The principle of RAG is to combine retrieval and generation into a unified framework. Similar feature selection techniques have shown robustness in dealing with complex datasets, as demonstrated by Shen et al~\cite{b31}.

\begin{figure}[ht!]
\centerline{\includegraphics[width=0.4\textwidth]{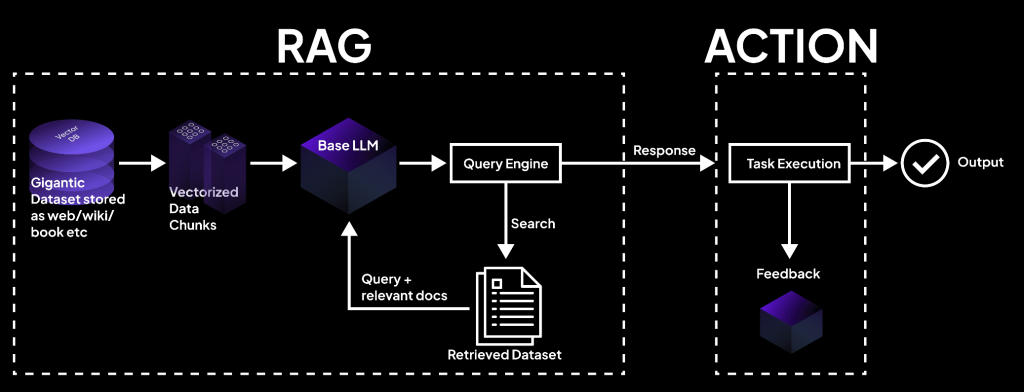}}
\caption{Architecture of RAG}
\label{fig:rag}
\end{figure}

RAG enhances response quality by combining retrieval and generative modeling. Given a query $q$, the retriever selects relevant documents $D = \{d_1, d_2, \ldots, d_N\}$ from a knowledge base. The retrieval step is represented as:

\[
P(d_i \mid q) = \frac{e(q) \cdot e(d_i)}{\|e(q)\|\|e(d_i)\|}
\]

where $e(q)$ and $e(d_i)$ are the embeddings of the query and document $d_i$. This is calculated using cosine similarity. The generator uses these retrieved documents to generate a response y: 

\[
P(y \mid q) = \sum_{d \in D} P(y \mid q, d) P(d \mid q)
\]

where $P(y|q,d)$ is the conditional probability of generating response y given query q and document d. 
	
OpenAI embeddings are vector representations capturing the semantic meaning of text. Ensemble methods have been shown to improve financial forecasting performance~\cite{b32}. Given input x, the embedding $e(x) \in R^d$ is generated by a pre-trained model. The similarity between two embeddings $e(x_1)$ and $e(x_2)$ is calculated as:

\[
sim(e(x_1), e(x_2)) = \frac{e(x_1) \cdot e(x_2)}{\|e(x_1)\| \|e(x_2)\|}
\]

where represents the dot product, and $\| e(x) \|$ is the Euclidean norm. 

ChromaDB is a vector database for managing high-dimensional embeddings. The retrieval process finds the embedding $e_j \in E$ that minimizes the distance to a query embedding $e(q)$:

\[
e^* = arg \min_{e_j \in E} \| e(q) - e_j \|
\]

where $\|e(q)-e_j\|$represents the distance (typically Euclidean or cosine) between the query embedding $e(q)$ and a candidate embedding $e_j$.

\subsection{Vulnerability Detection}
\begin{figure}[ht!]
\centerline{\includegraphics[width=0.3\textwidth]{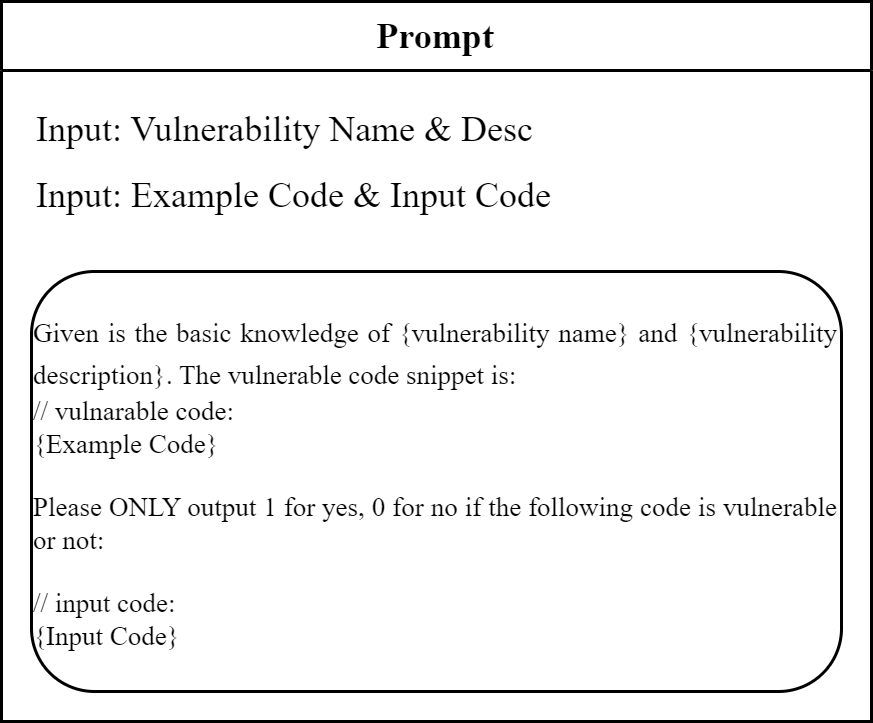}}
\caption{Prompt without CoT}
\label{fig:noCoT}
\end{figure}

\begin{figure}[ht!]
\centerline{\includegraphics[width=0.3\textwidth]{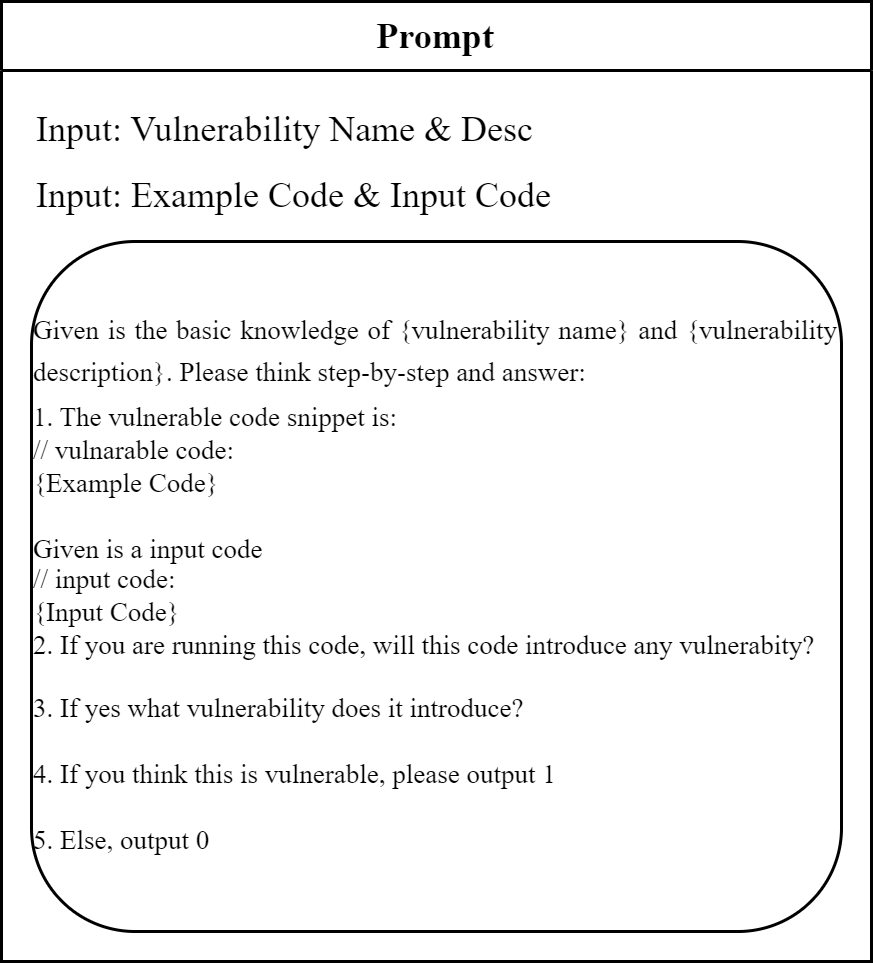}}
\caption{Prompt with CoT}
\label{fig:CoT}
\end{figure}
LProtector uses binary classification for vulnerability detection,Techniques that have also been reviewed in other AI/ML contexts~\cite{b33}. It was observed that GPT-4o identifies most input code snippets as vulnerable. Therefore, it is important to maintain balance in our test dataset to achieve unbiased detection results, which will be further investigated during the experiment section.

The OpenAI Embedding algorithm is used to transform every new input code snippet into word vectors. However, the revised embeddings are not saved directly in the database. Instead, they are compared with the existing vectors in the database in order to identify the top 5 most similar entries. GPT-4o later examines these records and selects the best one based on contextual similarities. The improved hint is created, as shown in Fig.~\ref{fig:noCoT}.

The input is put into GPT-4o, resulting in the final classification result. Basically, LProtector uses a vector database offline to perform semantic or pattern matching, ensuring that the achieved outcomes provide the necessary context for accurate vulnerability identification.

\section{Experiments}
\subsection{Experiments Procedure}
The Big-Vul dataset has a severe data imbalance problem~\cite{b22}, where the number of non-vulnerable samples significantly outweighs the vulnerable samples. Specifically, as shown in Table 1, the ratio of vulnerable to non-vulnerable samples is only 5.88\%.

\begin{table}[h]
    \centering
    \caption{Statistics of Big-Vul dataset}
    \begin{tabular}{|c|c|c|c|}
        \hline
        \multirow{2}{*}{\centering Dataset} & \multicolumn{3}{c|}{Data} \\ \cline{2-4} & Samples & Vul & Non-Vul \\ \hline
        Big-Vul  & 179,299  & 10,547 (5.88\%) & 168,752 (94.12\%) \\ \hline
    \end{tabular}
    \label{tab:big_vul_statistics}
\end{table}

In order to enhance our experiments, we initially conducted sampling to equalize the data, aiming for a 1:1 balance between vulnerable and non-vulnerable test data, which produced a total of 5,000 test instances. Next, we selected 500 vulnerable cases from the non-test data at random and collected the associated vulnerable code sections, which were saved in a vector database using RAG. Afterward, we conducted a comparison of the results by running LProtector together with VulDeePecker and Reveal. Similar designs are used to study the effects of visual field on thermal perception.~\cite{b34}

Then, to assess the internal workings of LProtector, we investigated how two important factors affected its performance:

\begin{enumerate}
    \item With/Without RAG:
    \par To investigate the effect of RAG on LProtector’s performance, we assess its results on the same dataset without the RAG component.
    \item With/Without CoT:
    \par We assessed the performance of LProtector on the same dataset without using CoT to understand its impact on vulnerability detection. 
\end{enumerate}

Then we need to evaluate the performance of LProtector with no RAG and CoT, which means purely LLM.

\subsection{Performance Metrics}\label{AA}
To evaluate the performance, We use Accuracy, Precision, Recall, F1~\cite{b6},~\cite{b35},~\cite{b36},~\cite{b37}.

\begin{enumerate}
    \item Accuracy: Accuracy is the most intuitive metric, representing the proportion of correctly classified instances (both vulnerabilities and non-vulnerabilities) to the total instances. It is calculated as the ratio of true positives (TP) and true negatives (TN) to the sum of all predicted classes, including false positives (FP) and false negatives (FN). Although accuracy is a useful metric, it can be misleading in the case of imbalanced datasets, where one class significantly outweighs the other, as is the case in our dataset.

\[
Accuracy = \frac{TP + TN}{TP + TN + FP + FN}
\]

    \item Precision: Precision measures the proportion of correctly predicted positive instances out of all instances predicted as positive. It is particularly important in cases where the cost of false positives is high. Precision is sensitive to the number of false positives, meaning that a model with high precision minimizes the occurrence of incorrectly classifying non-vulnerabilities as vulnerabilities.

\[
Precision = \frac{TP}{TP + FP}
\]

    \item  Recall: Recall (also known as sensitivity or true positive rate) quantifies the model’s ability to correctly identify actual positive instances. It is especially important in contexts where missing positive instances (false negatives) can have severe consequences. For our vulnerability detection task, recall reflects the model’s ability to correctly identify vulnerable code. 

\[
Recall = \frac{TP}{TP + FN}
\]

    \item FI Score: 	Finally, the F1 Score is the harmonic mean of Precision and Recall, providing a balanced measure of the two when there is a tradeoff between them. The F1 score is particularly useful when the dataset is imbalanced, as it provides a single metric that accounts for both false positives and false negatives. It offers a more nuanced measure of performance, especially when neither Precision nor Recall can fully capture the model’s efficacy in isolation.


\[
F_1 = 2 \times \frac{Precision \times Recall}{Precision + Recall}
\]

	By combining these metrics, we can assess the model’s performance across multiple dimensions, ensuring a balanced understanding of its classification capabilities. Each metric highlights different aspects of performance, providing insights into areas where the model excels and where improvements may be necessary.

\end{enumerate}

\subsection{Experiment Results}
In the initial experiments, we compared the performance of VulDeePecker, Reveal, and LProtector on the Big-Vul dataset. As shown in the table, LProtector outperformed the other two baselines across all metrics.

\begin{table}[h]
    \centering
    \caption{Overall Performance}
    \begin{tabular}{|c|c|c|c|c|}
        \hline
        \multirow{2}{*}{Baseline} & \multicolumn{4}{c|}{Metrics (\%)} \\ \cline{2-5}
                                   & Accuracy & Precision & Recall & F1 Score \\ \hline
        VulDeePecker              & 81.19    & 38.44     & 12.75  & 19.15    \\ \hline
        Reveal                    & 87.14    & 17.22     & 34.04  & 22.87    \\ \hline
        \textbf{LProtector}       & \textbf{89.68} & \textbf{30.52} & \textbf{38.07} & \textbf{33.49} \\ \hline
    \end{tabular}
    \label{tab:overall_performance}
\end{table}

LProtector achieved an accuracy of 89.68\%, which is higher than VulDeePecker’s 81.19\% and Reveal’s 87.14\%, indicating its stability in classifying both vulnerable and non-vulnerable samples. Although VulDeePecker had a higher precision (38.44\%) compared to LProtector (30.52\%), its recall was only 12.75\%, much lower than LProtector’s 38.07\%, meaning it struggled to capture true vulnerabilities. In contrast, LProtector effectively reduced false positives while identifying more real vulnerabilities. As a result, LProtector achieved an F1 score of 33.49\%, demonstrating a notable improvement over VulDeePecker’s 19.15\% and Reveal’s 22.87\%, showcasing its balanced performance in terms of precision and recall.

\begin{table}[h]
    \centering
    \caption{LProtector Variable Test}
    \begin{tabular}{|c|c|c|c|c|}
        \hline
        \multirow{2}{*}{Variables} & \multicolumn{4}{c|}{Metrics (\%)} \\ \cline{2-5}
                                   & Accuracy & Precision & Recall & F1 Score \\ \hline
        RAG + CoT                  & 89.68    & 30.52     & 38.07  & 33.49    \\ \hline
        No RAG                     & 76.42    & 22.71     & 25.30  & 23.92    \\ \hline
        No CoT                     & 79.73    & 24.85     & 28.40  & 26.51    \\ \hline
        No RAG \& CoT              & 68.19    & 17.85     & 19.42  & 18.61    \\ \hline
    \end{tabular}
    \label{tab:lprotector_variable_test}
\end{table}

Removing the RAG component led to a significant drop in LProtector’s performance. The accuracy decreased to 76.42\%, which is lower than VulDeePecker’s 81.19\%, and the F1 score dropped to 23.92\%. This result suggests that RAG plays a critical role in retrieving relevant context for accurate vulnerability detection. Without RAG, LProtector cannot get access to prior knowledge, which directly impacts its ability to correctly identify vulnerabilities.

Removing CoT causes a drop in performance, resulting in accuracy decreasing to 79.73\% and the F1 score decreasing to 26.51\%. This falls just under the baseline accuracy of 81.19\%, suggesting that CoT improves the model's reasoning skills. The accuracy of LProtector's forecasts decreases without CoT, impacting precision and recall.

When both RAG and CoT were removed, LProtector’s accuracy fell sharply to 68.19\%, and its F1 score was reduced to 18.61\%. This performance is significantly lower than all other configurations, highlighting that RAG and CoT are both essential for LProtector’s effectiveness. Without these components, the model struggles to process and interpret the code, making it almost ineffective in detecting vulnerabilities.

In the two experiments, we initially showed that LProtector surpasses VulDeePecker and Reveal on the Big-Vul dataset, obtaining better F1 scores and overall performance. Next, we examined the inner workings by disassembling the RAG and CoT parts individually. The findings indicated that the model was most affected by RAG, as its removal caused a notable decrease in both accuracy and F1 score. Getting rid of CoT had a minor impact, but still led to performance dropping below the original level. LProtector's performance significantly decreased to its lowest level when RAG and CoT were both eliminated, underscoring the essential nature of these two components for successful vulnerability detection.

\section{Conclusion}
In the latest studies on utilizing LLMs for identifying vulnerabilities, code is often viewed as regular text that is fed directly into the model, without the necessary contextual information and domain expertise needed for precise detection. In response to these restrictions, we suggest LProtector, a new vulnerability detection system using advanced retrieval methods and Chain of Thought (CoT) reasoning to improve prediction accuracy. LProtector uses RAG to insert expertise specific to domains, enhancing its comprehension of intricate code structures. The Big-Vul dataset experiment showed that LProtector outperforms other baseline methods with a 33.49\% F1 score, surpassing VulDeePecker and Reveal. Additionally, analysis of components shows that both RAG and CoT are essential for the model's success, with RAG playing the biggest role in enhancing prediction accuracy. In our future efforts, we aim to improve the retrieval and reasoning methods of LProtector to increase its ability to detect threats. Furthermore, we will investigate the potential of utilizing LProtector in various intricate software systems, including mobile apps and cloud environments, to assess its capability to be applied broadly. Ultimately, our goal is to explore combining automated vulnerability repair methods with LProtector, which presents a difficult but promising avenue for future studies.

\section*{Acknowledgment}

We would like to express our gratitude to the various institutions for their valuable contributions to data collection. Additionally, we extend our appreciation to all members of our research team for their dedication and collaborative efforts throughout the development of LProtector.

\end{document}